\documentclass[a4paper,11pt]{article}
\usepackage[utf8]{inputenc}
\usepackage[margin=1in]{geometry}

\usepackage{amsmath,amssymb,amsfonts,amsthm}
\usepackage{graphicx,tikz,color,textcomp}
\usepackage{hyperref}

\usepackage{epsfig}
\usepackage{array}
\usepackage{lineno}

\newcommand{\ket}[1]{|#1\rangle}
\newcommand{\braket}[2]{\langle #1|#2\rangle}

\newcommand{\THM}[2]{\vspace*{12pt} \noindent {\bf Theorem~#1:} #2 }
\newcommand{\MTHM}[1]{\vspace*{12pt} \noindent {\bf Main Theorem:} #1 }
\newcommand{\CORR}[2]{\vspace*{12pt} \noindent {\bf Corollary~#1:} #2 }
\newcommand{\PROOF}[1]{\vspace*{12pt} \noindent {\bf Proof:} #1 $\square\,$}

\newcommand{\mymatrix}[2]{\left( \begin{array}{#1} #2 \end{array} \right)}
\newcommand{\myvector}[1]{\mymatrix{c}{#1}}
\newcommand{\mypar}[1]{\left( #1 \right)}
\newcommand{\myceil}[1]{ \left \lceil #1 \right\rceil }

\title{Quantum Logarithmic Space and Post-selection}
\author{Fran{\c c}ois Le Gall$^1$ \and Harumichi Nishimura$^2$ \and Abuzer Yakary\i lmaz$^{3,4}$}
\date{
\small
$^1$Graduate School of Mathematics, Nagoya University, Nagoya, Japan
\\
\small
$^2$Graduate School of Informatics, Nagoya University, Nagoya, Japan
\\
\small
$^3$Center for Quantum Computer Science, University of Latvia, R\=\i ga, Latvia
\\
\small
$^4$QWorld Association, https://qworld.net
\\
\small Emails: \textit{legall@math.nagoya-u.ac.jp, hnishimura@i.nagoya-u.ac.jp, abuzer@lu.lv}
}

\begin{document}

\maketitle

\begin{abstract}
    Post-selection, the power of discarding all runs of a computation in which an undesirable event occurs, is an influential concept introduced to the field of quantum complexity theory by Aaronson (Proceedings of the Royal Society A, 2005).
	In the present paper, we initiate the study of post-selection for space-bounded quantum complexity classes. Our main result shows the identity {$\sf PostBQL=PL$}, i.e., the class of problems that can be solved by a bounded-error (polynomial-time) logarithmic-space quantum algorithm with post-selection ({$\sf PostBQL$}) is equal to the class of problems that can be solved by unbounded-error logarithmic-space classical algorithms ({$\sf PL$}). This result gives a space-bounded version of the well-known result {$\sf PostBQP=PP$} proved by Aaronson for polynomial-time quantum computation.
	As a by-product, we also show that $\sf PL$ coincides with the class of problems that can be solved by bounded-error logarithmic-space quantum algorithms that have no time bound.
\end{abstract}

\section{Introduction}\label{sec:introduction}\vspace{-2mm}
\paragraph{Post-selection.}
Post-selection is the power of discarding all runs of a computation in which an undesirable event occurs. This concept was introduced to the field of quantum complexity theory by Aaronson~\cite{Aar05}. While unrealistic, post-selection turned out to be an extremely useful tool to obtain new and simpler proofs of major results about classical computation, and also prove new results about quantum complexity classes. The most celebrated result is arguably the identity {$\sf PostBQP =PP$} proved by Aaronson \cite{Aar05}, which shows that the class of problems that can be solved by a bounded-error polynomial-time quantum algorithm with post-selection ({$\sf PostBQP$}) is equal to the class of problems that can be solved by unbounded-error polynomial-time classical algorithms ({$\sf PP$}), and thus makes possible to bridge quantum complexity classes and classical complexity classes.\vspace{-4mm}

\paragraph{Space-bounded quantum complexity classes.}
The study of space-bounded quantum Turing machines was initiated by Watrous \cite{Wat99A}. Watrous showed in particular that any quantum Turing machine running in space $s$ can be simulated by an unbounded-error probabilistic Turing machine running in space $O(s)$. This result implies the identity {$\sf PQL=PL$}, where {$\sf PQL$} denotes the class of problems that can be solved by unbounded-error logarithmic-space quantum Turing machines, and {$\sf PL$} denotes the class of problems that can be solved by unbounded-error logarithmic-space classical Turing machines. The main open question of the field is whether bounded-error quantum Turing machines can be simulated space-efficiently by \emph{bounded-error} classical Turing machines.

A major step towards establishing the superiority of space-bounded quantum Turing machines over  space-bounded classical (bounded-error) Turing machines has been the construction by Ta-Shma \cite{TaS13} of 
logarithmic-space quantum algorithms for inverting well-conditioned matrices (it is unknown how to perform the same task classically in logarithmic space). While Ta-Shma's quantum algorithm used intermediate measurements, a version of this quantum algorithm without measurement was later constructed by Fefferman and Lin \cite{FL16A} (see also \cite{FKLMN16} for a related result on space-efficient error reduction for unitary quantum computation). Very recent works \cite{FR21,GRZ20} have further showed that many other problems from linear algebra involving well-conditioned matrices can be solved as well in logarithmic space by quantum algorithms, and additionally showed that intermediate measurements can be removed from any space-bounded quantum computation.\vspace{-4mm}

\paragraph{Our results.}
In view of the impact of the concept of post-selection to quantum complexity theory and in view of the surge of recent activities on space-bounded quantum complexity classes, a natural question is investigating the power of post-selection for space-bounded quantum complexity classes. To our knowledge, this question has not been investigated so far in the literature (while the notion of post-selection was previously studied in quantum automata theory~\cite{YS13A}). In this paper, we tackle this question and obtain the following result (here {$\sf PostBQL$} denotes the class of problems that can be solved by a bounded-error polynomial-time logarithmic-space quantum Turing machine that uses post-selection --- see Section \ref{sec:prelim} for a formal definition):

\MTHM{$\sf  PostBQL = PL$.}\vspace{3mm}

This result thus gives a space-bounded version of the result {$\sf PostBQP=PP$} mentioned above for polynomial-time complexity classes. This enables us to bridge quantum complexity classes and classical complexity classes for space-bounded computation as well, and thus suggests that post-selection may become a useful tool to analyze space-bounded (quantum and classical) computation as well. 
Actually, as a by-product of our main result, we also obtain the fact that $\sf PL$  coincides with the class of problems that can be solved by bounded-error logarithmic-space quantum algorithms that has no time bound (namely, the bounded-error logarithmic-space quantum algorithms are as computationally powerful as the unbounded-error ones under no time restriction).

We additionally present several results about logarithmic-space quantum computation with post-selection in Section \ref{sec:consequences}.\vspace{-4mm}

\paragraph{Overview of our techniques.}
As for the result {$\sf PostBQP=PP$} proved by Aaronson \cite{Aar05}, the nontrivial part of the proof of our main theorem is the simulation of a probabilistic machine by a post-selecting quantum simulation machine. The simulation technique given in \cite{Aar05} requires a polynomial amount of qubits, and thus cannot be used in our setting since we are limited to a logarithmic amount of qubits. Therefore, we propose a different simulation, which is composed of three parts. 
First, we show how to simulate the computation of a logarithmic-space probabilistic Turing machine by a logarithmic-width probabilistic circuit $K$ (Section~\ref{sub:part1}). Note that the computation process of $K$ is represented by a mixture $\sum_j p_j C_j$, which means that the configuration is in $C_j$ with probability $p_j$. (It can be written as $\sum_j p_j |C_j\rangle\langle C_j|$ when the mixed state formalism~\cite{NC00} is used.) Here, we can assume that there are unique accepting and rejecting configurations $C_{a}$ and $C_{r}$. Thus, the final mixture of $K$ can be represented in the form of $p C_{a}+(1-p) C_{r}$, where $p>1/2$ if the input is a yes-instance, and $p<1/2$ if it is a no-instance. Second, 
we give a simulation of the probabilistic circuit $K$ by a logarithmic-space quantum Turing machine $M$ with post-selection (Section~\ref{sub:part2}). Note that this simulation is done in a {\em coherent} manner. Namely, if the mixture of $K$ at some step is $\sum_j p_j C_j$, the quantum state of $M$ at the corresponding simulation step should be the normalized state of $\sum_j p_j |C_j\rangle$. Thus, $M$ produces the normalized state of $|\psi\rangle=p|C_a\rangle+(1-p)|C_r\rangle$ as the final outcome. In fact, we use the power of post-selection for this simulation, and the final outcome can be obtained after post-selection with an exponentially small probability. Then, the third part is fairly similar to the approach used in~\cite{Aar05}: using polynomial number of states constructed from the same number of copies of $|\psi\rangle$, we use repetition and post-selection to increase the success probability of the simulation (Section~\ref{sub:part3}). 
 
\section{Preliminaries}\label{sec:prelim}


\subsection{Space-bounded probabilistic Turing machines}
\label{sec:PTM}

A classical space-bounded Turing machine has an input tape and a work tape. Both tapes are infinite and their cells are indexed by integers, each of which contains the blank symbol $ (\#) $ unless it is overwritten with a different symbol. The input tape has a read-only head and the work tape has a read/write head. Each head can access a single cell in each time step and, after each transition, it can stay on the same cell, move one cell to the right, or move one cell to the left.
 
The input alphabet is denoted $ \Sigma $ and the work tape alphabet is denoted $ \Gamma $, none of which contains the blank symbol. Moreover, $ \tilde{\Sigma} = \Sigma \cup \{\#\} $ and $ \tilde{\Gamma} = \Gamma \cup \{\#\} $. For a given string $ x $, $ |x| $  represents the length of $ x $. 

Formally, a (space-bounded) probabilistic Turing machine (PTM) $ M $ is a 7-tuple
\[
	M = (S,\Sigma,\Gamma,\delta,s_i,s_a,s_r),
\]
where $ S $ is the set of (internal) states, $ s_i \in S $ is the initial state, $ s_a \in S $ and $ s_r \in S $ ($s_a \neq s_r $) are the accepting and rejecting states, respectively, and $ \delta $ is the transition function described below.

At the beginning of the computation, the given input, say $ x \in  \Sigma^* $, is placed on the input tape between the first cell and the $ |x| $-th cell, the input tape head and the work tape head are placed on the cells indexed by 0s, and the state is set to $ s_i $. In each step, $ M $ evolves with respect to the transition function and the computation is terminated after entering $ s_a $ or $ s_r $. In the former (latter) case, the decision of ``acceptance'' (``rejection'') is made. It must be guaranteed that the input tape head never visits the cells indexed by $ -1 $ and $ |x|+2 $. The formal definition of $ \delta $ is as follows:
\[
	\delta: S \times \tilde{\Sigma} \times \tilde{\Gamma} \times S \times \tilde{\Gamma} \times \{-1,0,1\} \times \{ -1,0,1 \} \rightarrow \left\{0,\frac{1}{2},1\right\}.
\]
Suppose that $ M $ is in $ s \in S $ and reads $ \sigma \in \tilde{\Sigma} $ and $ \gamma \in \tilde{\Gamma} $ on the input and work tapes, respectively. Then, in one step, the new state is set to $ s' \in S $, the symbol $ \gamma' \in \tilde{\Gamma} $ is written on the cell under the work tape head, and the positions of the input and work tape heads are respectively updated with respect to $ d_i \in \{-1,0,1\} $ and $ d_w \in \{-1,0,1\} $, with probability
\[
	\delta(s,\sigma,\gamma,s',\gamma',d_i,d_w),
\]
where the input (work) tape head moves one cell to the left if $ d_i = -1 $ ($ d_w = -1 $) and one cell to the right if $ d_i =1 $ ($d_w = 1$). Remark that any transition with zero probability is never implemented. To be a well-formed PTM, for each triple $ (s,\sigma,\gamma) $, 
\[
	\sum_{s' \in S, \gamma' \in \tilde{\Gamma},d_i \in \{-1,0,1\},d_w \in \{-1,0,1\}} \delta(s,\sigma,\gamma,s',\gamma',d_i,d_w) = 1.
\]

For a given input $ x \in \Sigma^* $, $ M $ can follow more than one computation path. A computation path either halts with a decision or runs forever. A halting path is called accepting (rejecting) if the decision of ``acceptance'' (``rejection'') is made on this path. The accepting (rejecting) probability of $ M $ on $ x $ is the cumulative sum over all accepting (rejecting) paths. 

A language $  L $ is said to be recognized by PTM $ M $ with unbounded error if and only if any $ x \in L $ is accepted by $ M $ with probability more than $ 1/2 $ and any $ x \notin L $ is accepted with probability less than $ 1/2 $. 
A language $ L $ is said to be recognized by PTM $ M $ with error bound $ \varepsilon < 1/2 $ if and only if any $ x \in L $ is accepted by $ M $ with probability at least $ 1 - \varepsilon $ and any $ x \notin L $ is rejected with probability at least $ 1 - \varepsilon $. When $\varepsilon >0$ is a constant (independent of the input), it is said that $ L $ is recognized by $ M $ with bounded error. As a special case, if all non-members of $L$ are accepted with probability 0, then it is called one-sided bounded-error. 
A PTM making only deterministic transitions (i.e., such that the range of the transition function is $\{0,1\}$) is a deterministic Turing machine (DTM). 

The range of the transition function can also be defined as $ [0,1] \cap \mathbb{Q} $, and thus the PTM, called rational valued PTM, can make more than one transition with rational valued probabilities in each step. Remark that all results presented in this paper are also followed for rational valued PTMs. A nondeterministic Turing machine (NTM) can be defined as a rational valued PTM and a language is said to be recognized by a NTM if and only if for any member there is at least one accepting path and for any non-member there is no accepting path (or equivalently any member is accepted with nonzero probability and any non-member is accepted with zero probability). 

A language is recognized by a machine in (expected) time $ t(n) $ and space $ s(n) $ if the machine, on a given input $ x $, runs no more than (expected) $ t(|x|) $ time steps and visits no more than $ s(|x|) $ different cells on its work tape with non-zero probability. 

The class $ \mathsf{PL}$ ($\mathsf{L}$ and $\mathsf{NL}$) is the set of languages recognized by unbounded-error PTMs (DTMs and NTMs) in logarithmic space (with no time restriction). It is shown that each of these classes coincides with the subclass such that the running time of the corresponding machines is polynomially bounded (note that the proof is nontrivial for $\mathsf{PL}$~\cite{Jung85}).

The class $\mathsf{BPL} $ ($ \sf RL $) is the set of languages recognized by bounded-error PTMs (one-sided bounded-error PTMs) {\em in polynomial time} 
and logarithmic space. 
On contrary to the above three classes $\sf PL, L, NL$, 
it is unknown that these two classes are the same as their corresponding classes such that the underlying machines have no time restriction, which we denote by $\sf BPL(\infty)$ ($\sf RL(\infty)$). 

Any language $ L $ is in $ \sf C_{=}L $ \cite{AO96} if and only if there exists a polynomial-time logarithmic-space PTM $ M $ such that any $ x \in L $ is accepted by $ M $ with probability $ \frac{1}{2} $ and any $ x \notin L $ is accepted by $ M $ with probability other than $ \frac{1}{2} $. 

\subsection{Turing machines with post-selection}

A postselecting PTM (PostPTM) has the ability to discard some predetermined outcomes and then makes its decision with the rest of the outcomes, which is guaranteed to happen with non-zero probability (see \cite{Aar05,YS13A}). Formally, a PostPTM is a modified PTM with three halting states. A PTM has the accepting state $ s_a $ and the rejecting state $ s_r $ as the halting states. A PostPTM has an additional 
halting state $ s_n $ called the non-postselecting halting state. In this paper, we require that a PostPTM must halt its computation absolutely, i.e., there is no infinite loop. 

For a given input $ x $, let $ p_{acc,M}(x) $ ($ p_{rej,M}(x) $ and $ p_{npost,M}(x) $) be the probability of PostPTM $ M $ ending in $ s_a $ ($s_r$ and $s_n$). Since $ M $ halts absolutely, we know that 
\[
	p_{acc,M}(x) + p_{rej,M}(x) + p_{npost,M}(x) = 1.
\]
Due to post-selection, we discard the probability $ p_{npost,M}(x) $ and then make a normalization on $ p_{acc,M}(x) $ and  $ p_{rej,M}(x) $ for the final decision. Thus, the input $ x $ is accepted (rejected) by $ M $ with probability
\[
	\tilde{p}_{acc,M}(x):=\frac{p_{acc,M}(x)}{ p_{acc,M}(x) + p_{rej,M}(x) } ~~~ \left(\tilde{p}_{rej,M}(x):=\frac{p_{rej,M}(x)}{ p_{acc,M}(x) + p_{rej,M}(x) }\right).
\]

The postselecting counterparts of $ \mathsf{BPL} $ and $\sf RL$ are $ \mathsf{PostBPL} $ and $ \sf PostRL $, respectively. (For instance, $L$ is in $\sf PostBPL$ if and only if there are a polynomial-time logarithmic-space PostPTM $M$ and a constant $\varepsilon<1/2$ such that $\tilde{p}_{acc,M}(x)$ is at least $1-\varepsilon$ when $x$ is in $L$, and $\tilde{p}_{rej,M}(x)$ is at least $1-\varepsilon$ when $x$ is not in $L$).
Let $\sf PostEPL$ denote the class of languages recognized with no error (or exactly) by polynomial-time logarithmic-space PostPTMs (i.e., $L$ is in $\sf PostEPL$ if and only if there is a polynomial-time logarithmic-space PostPTM $M$ such that $p_{acc,M}(x)>0$ and $p_{rej,M}(x)=0$ when $x$ is in $L$, and $p_{acc,M}(x)=0$ and $p_{rej,M}(x)>0$ when $x$ is not in $L$).

\subsection{Space-bounded quantum Turing machines}

The initial quantum Turing machine (QTM) models (e.g., \cite{Deu85,BV97,Wat99A}) were defined fully quantum. 
While quantum circuits have been used more widely in literature, QTMs are still the main computational models 
when investigating space bounded complexity classes. However, their definitions have been modified since 90s (e.g., \cite{Wat03,MW12,TaS13,FL16A}). The main modifications are that the computation is governed classically and the quantum part can be seen like a quantum circuit. This paper follows these modifications. To be more precise, our QTM is a PTM augmented with 
a quantum tape. Here, the quantum tape is designed like a quantum circuit, i.e., it contains a qubit (or qudit) in each tape cell and it can have more than one tape head so that a quantum gate can be applied to a few qubits at the same time. 

We remark that the result given in this paper can also be obtained by the other space-bounded QTMs defined in literature \cite{Wat09A,Wat03,YS11A,MW12,TaS13}, where algebraic numbers are used as transition values. The main advantage of the aforementioned modifications in QTMs is to simplify the proofs and the descriptions of quantum algorithms.

Formally, a (space-bounded) QTM $ M $ is a 9-tuple
\[
	M = (S,\Sigma,\Gamma,\delta_q,\delta_c,s_i,s_a,s_r,\Omega),
\]
where, different from the PTMs, the transition function is composed by two functions $ \delta_q $ and $ \delta_c $ that are responsible for the transitions on quantum and classical parts, respectively, and $ \Omega $ is the set of contents of a classical register storing quantum measurement outcomes. (Similarly to the PTMs, $S$ is the set of internal states, $\Sigma$ is the input alphabet, $\Gamma$ is the work tape alphabet, and $s_i$, $s_a$, and $s_r$ are respectively the initial state, the accepting state, and the rejecting state.) As the physical structure, $M$ additionally has a quantum tape with $ l $ heads, and the classical register storing a value in $ \Omega=\{1,\ldots,m\} $, where $ l,m>0 $ are constants (independent of the input given to $M$). The quantum tape heads are numbered from 1 to $ l $. 
For simplicity, we assume that the quantum tape contains only qubits (with states $ \ket{0} $ and $ \ket{1} $) in its cells. Each cell is set to $ \ket{0} $ at the beginning of the computation. For a given input $ x \in \Sigma^* $, the classical part is initialized as described for PTMs. The $ l $ tape heads on the quantum tape are placed on the qubits numbered from 0 to $ l-1 $. 

The overall computation of $ M $ is governed classically. Each transition of $ M $ has two phases, quantum and classical, which alternate. We define the transition functions $ \delta_q $ and $ \delta_c $ different from the transition functions of  PTMs. Suppose that $ M $ is in $ s \in S $ and reads $ \sigma \in \tilde{\Sigma} $ and $ \gamma \in \tilde{\Gamma} $, respectively. For each triple $ (s,\sigma,\gamma) $, $ \delta_q (s,\sigma,\gamma)$ can be the identity operator, a projective measurement (in the computational basis), or a unitary operator. If it is the identity operator, the quantum phase is skipped by setting the value of the classical register to 1 (in $\Omega$). 
If the quantum operator is unitary, then the corresponding unitary operator is applied to the qubits under the heads on the quantum tape, and the value in the classical register is set to 1 (in $\Omega$). If it is a measurement operator, then the corresponding projective measurement is done on the qubits under the heads on the quantum tape, and the measurement outcome, represented by an integer between 1 and $ m' \leq m $ (in $\Omega$), is written in the classical register, where $m'$ is the total number of all possible measurement outcomes of the measurement operator. 

After the quantum phase, the classical phase is implemented. For each quadruple $( s, \sigma,\gamma, \omega )$, $ \delta_c $ returns the new state, the symbol written on the work tape, and updates of all heads, where $ \omega \in \Omega $. 

The termination of the computation of $M$ is the same as the PTMs, i.e., done by entering the accepting state $s_a$ or the rejecting state $s_r$. One time step corresponds to a single transition. We add the number of qubits visited with non-zero probability during the computation (as well as the number of cells visited on the classical work tape) to the space usage.

Remark that any QTM using superoperators can be simulated by a QTM using unitary operators and measurements with negligible memory and time overheads, i.e., by using extra quantum and classical states, any superoperator can be implemented by unitary operators and measurements in constant steps (e.g. \cite{NC00,SayY14}).

Since the computation of the QTM defined above is controlled classically, a postselecting QTM (PostQTM) can be defined similar to PostPTMs: the PostQTM has an additional classical halting state $ s_n $, and any computation that ends in $ s_n $ is discarded when calculating the overall accepting and rejecting probability on the given input.

The quantum counterparts of $ \sf BPL $ ($\sf BPL(\infty)$), $\sf RL$, $ \sf PL $, $ \sf NL $, $ \sf PostBPL $, $\sf PostRL$, and $ \sf PostEPL $ are $ \sf BQL $ ($\sf BQL(\infty)$), $\sf RQL$, $ \sf PQL $, $ \sf NQL $\footnote{Note that $\sf NQL$ is the quantum counterpart of $\sf NL$ based on the criterion by the accepting probabilities of the underlying machine, not the certificate-based counterpart ($\sf QMAL$).}, $ \sf PostBQL $, $\sf PostRQL$, and $ \sf PostEQL $, respectively, where QTMs use algebraic numbers as transition amplitudes. 

The following relations on logarithmic space quantum and classical complexity classes are already known \cite{Imm88,Wat99A,Wat03,FR21}:
\[
	\sf
	L \subseteq NL = coNL \subseteq coC_=L = NQL \subseteq PL = PQL.
\]
\[
	\sf
	L \subseteq BPL \subseteq BQL 
	\subseteq PL = PQL.
\]

\section{Main Result}\label{section:main}

In this section, our main theorem ($\sf PostBQL=PL$) is proved. We start with the easy inclusion.

\THM{1}{$\sf  PostBQL \subseteq PQL = PL  $}

\PROOF{Any polynomial-time logarithmic-space PostQTM $ M $ can be easily converted to a polynomial-time logarithmic-space QTM $ M' $ such that  $ M' $ enters the accepting and rejecting states with equal probability when $ M $ enters the non-postselecting halting state. Thus the balance between accepting and rejecting probabilities is preserved, and thus the language recognized by $ M $ with bounded-error is recognized by $ M' $ with unbounded error.
}

In the rest of this section, we give the proof of the following inclusion.

\THM{2}{ $ \sf PL \subseteq PostBQL $.}
\\

As described in Section~\ref{sec:introduction}, the proof of Theorem 2 consists of three parts, each of which will be given in the next three subsections. 
We start by giving an overview of the first part. Let $ L $ be a language in $ \sf PL $. Then there exists a PTM $ M $ recognizing $ L $ with unbounded error such that $ M $ on input $ x $ halts in $ |x|^{k} $ steps by using at most $ d \log(|x|) $ space for some fixed positive integers $ d $ and $ k $. 

Without loss of generality, we can assume that $ M $ always splits into two paths in every step, the work tape alphabet of $ M $ has only two symbols $ 0 $ and $ 1 $, and $ M $ halts only when the work tape contains only blanks and both tape heads are placed on the 0-th cells, i.e., there exist a single accepting and a single rejecting configurations. Let $ m $ be the number of internal states.

We fix $ x $ as the given input with length $ |x| = n $. Any configuration of $ M $ is represented by a 4-tuple of binary strings
\[
	(s,h_{\rm in},w,h_{\rm wk}),
\]
where $ s $ is the internal state, $ h_{\rm in} $ is the position of the input head, $ w $ is the content of the work tape, and $ h_{\rm wk} $ is the position of the work tape. (We also assume that $ w $ is always a binary string, which does not contain any blank symbol.) The set of all configurations is denoted by $ C^x $, i.e., $ C^x = \{ C_1,\ldots,C_N \} $ for some $ N $ polynomial in $ n $. The length of any configuration is 
\[
	l = \myceil{\log m} + \myceil{ \log n } + \myceil{ d \log n }+ \myceil{ \log (d \log n) } \in O(\log n) .	
\]

Based on $ C^x $, we define a stochastic matrix $ P_x $, called \textit{the configuration matrix}, whose columns and rows are indexed by configurations and its $ (j,i) $-th entry represents the probability going from $ C_i $ to $ C_j $. Then, the whole computation of $ M $ on $ x $ can be traced by an $ N $-dimensional column vector, called \textit{configuration vector}:
\[
	v_{l+1} =  P_x v_l,
\]
where $ 1 \leq l \leq n^k  $ and $ v_l $ represents the probability distribution of the configurations after the $ l $-th step. Here, $ v_0 $ is the initial configuration vector having a single nonzero entry, that is 1, corresponding to the initial configuration, and $ v_{n^k} $ is the final configuration vector having at most two nonzero entries that keep the overall accepting and rejecting probabilities:
\[
	v_{n^k} = P_x^{n^k} v_0.
\] 
Since the computation is split into two paths in each step with equal probability, the overall accepting ($ A $) and rejecting probabilities ($R$) are respectively of the forms 
\[
	\frac{A'}{2^{n^k}} \mbox{ and } \frac{R'}{2^{n^k}},
\]
where $ 0 \leq A',R' \leq 2^{n^k} $, $ A'+ R' = 2^{n^k}$, and $ A' \neq 2^{n^k-1} $.

We present a simulation of the above matrix-vector multiplication in logarithmic space. It is clear that keeping all entries of a single configuration vector separately requires polynomial space in $ n $. On the other hand, a single configuration can be kept in logarithmic space. Therefore, we keep a mixture of configurations as a single summation for any time step. In other words, we can keep $ v_i $ as
\[
	v_i[1] C_1 + v_i[2] C_2 + \cdots + v_i[n^k] C_{n^k},
\] 
where each coefficient $v_i[j]$ represents the probability of being in the corresponding configuration $C_j$. The transition from $ v_i $ to $ v_{i+1} $ can be obtained in a single step by applying $ P_x $. However, in our simulation, we can do this in $ n^k $ sub-steps. The idea is as follows: In the $ j $-th sub-step, we check whether our mixture has $ C_j $ or not. If it exists, then $ C_j $ is evolved to $ C_{j}' $ and $ C_{j}'' $ that are the configurations obtained from $ C_j $ in a single step when the outcome of the coin is respectively heads or tails. In this way, from the mixture corresponding to $ v_i $, we obtain the next mixture:
\[
	v_{i+1}[1] C_1 + v_{i+1}[2] C_2 + \cdots + v_{i+1}[n^k] C_{n^k}.
\]
Then, the final mixture is
\[
	A  C_a + R C_r,
\]
where $ C_a $ and $ C_r $ are the accepting and rejecting configurations, respectively. 

We present the details of this simulation in the following subsection.

\subsection{Probabilistic circuit}\label{sub:part1}

In this subsection, it is shown that we can construct, in deterministic logarithmic space, a logarithmic-width and polynomial-depth probabilistic circuit $K_{M,x}$ that simulates $ M $ on $ x $.

Note that a logarithmic-space DTM can easily output each element of $ C^x $. Moreover, for any $ C_j \in C^x $, it can also easily output two possible next configurations $ C_j' $ and $ C_j'' $ such that $ M $ switches from $ C_j $ to $ C_j' $ if the result of the coin flip is heads and it switches from $ C_j $ to $ C_j'' $ if the result of the coin flip is tails. 

A logarithmic-space DTM 
$D$ described below can output the desired probabilistic circuit $K_{M,x}$ with width $ l+3 $ where (i) the first bit is named as \textit{the random bit} that is used for coin flip, (ii) the second and third bits are named as \textit{the block control bit} and \textit{the configuration control bit} that are used to control the transition between the configurations in each time step, and (iii) the rest of the bits hold a configuration of $ M $ on $ x $. 

The circuit $K_{M,x}$ consists of $n^k$ \textit{blocks}, and $ D $ outputs the $ n^k $ blocks. Each block corresponds to a single time step of $ M $ on $ x $:
\[
	block_1, block_2, \ldots, block_{n^k},
\]
where each block is identical, i.e., each block implements the transition matrix $ P_x $ operating on configurations. Remark that, after $ block_i $, we have the mixture representing $ v_i $. 

Before each block, {the random bit} is set to 0 or 1 with equal probability and {the block control bit} is set to 1. As long as the block control bit is 1, the configurations are checked one by one in the block. Once it is set to 0, the remaining configurations are skipped.

Any block is composed by $ N $ \textit{parts} where each part corresponds to a single configuration:
\[
	part_1,part_2,\ldots,part_N.	
\]
Here, $ part_j $ implements  the transitions from $ C_j $ in a single step. In $ part_j $, we do the following items:
\begin{enumerate}
	\item If \textit{the block control bit} is 0, then SKIP the remaining items. Otherwise, CONTINUE.
	\item SET \textit{the configuration control bit} to 1 (here we assume that $ M $ is in $ C_j $). 
	\item It checks whether $ M $ is in $ C_j $. 
    \begin{itemize}
    \item If $ M $ is not in $ C_j $, 
    SET \textit{the configuration control bit} to 0 and SKIP the remaining items. (Remark that the block control bit is still 1 in this case, and thus the next configuration $C_{j+1}$ will be checked in $part_{j+1}$.)
    \item Otherwise (i.e., if $ M $ indeed is in $ C_j $), CONTINUE. 
    \end{itemize}  
	\item SWITCH from $ C_j $ to $ C_j' $ if \textit{the random bit} is 0 and SWITCH from $ C_j $ to $ C_j'' $ if \textit{the random bit} is 1.
	\item SET \textit{the block control bit} to 0.
\end{enumerate}

After all $n^k$ blocks, $ D $ outputs the last block called \textit{decision block}. In the last block, it is checked whether the last configuration is $ C_a $ or $ C_r $. If it is $ C_a $ (resp.~$C_r$), then the first bit of the decision block is set to 1 (resp.~0).

\begin{figure}[h]
\centering
\fbox{
\begin{minipage}{0.85\textwidth}
\centering
\begin{tikzpicture}
\foreach \i in {0,...,2} {
\draw[thick, -] (0,5-0.5*\i) -- (4,5-0.5*\i);
\draw[thick, -] (5,5-0.5*\i) -- (7,5-0.5*\i);
\draw[thick, -] (8,5-0.5*\i) -- (10,5-0.5*\i);
}
\node at (0.75,5.14) {\footnotesize random bit};
\node at (1.15,4.64) {\footnotesize block control bit};
\node at (1.65,4.14) {\footnotesize configuration control bit};
\draw[thick, -] (0,3) -- (10,3);
\node at (1.5,2.8) {$ \vdots $};
\draw[thick, -] (0,2.4) -- (3,2.4) -- (4,3.5);
\draw[thick, -] (5,3.5) -- (6,2.4) -- (10,2.4);
\draw[thick, -] (0,1.9) -- (6,1.9) -- (7,3.5);
\draw[thick, -] (8,3.5) -- (9,1.9) -- (10,1.9);
\node at (1.5,1.7) {$ \vdots $};
\draw[thick, -] (0,1.3) -- (10,1.3);
\node at (0.2,3.2) {\footnotesize 1st};
\node at (0.3,2.6) {\footnotesize $j$-th};
\node at (0.55,2.1) {\footnotesize ($j$+1)-th};
\node at (0.3,1.5) {\footnotesize $l$-th};
\draw (4,5.2) -- (5,5.2) -- (5,3.5) -- (4,3.5) -- (4,5.2);
\node at (4.5,4.2) {$ H_j $};
\draw (7,5.2) -- (8,5.2) -- (8,3.5) -- (7,3.5) -- (7,5.2);
\node at (7.5,4.2) {$ H_{j+1} $};
\end{tikzpicture}
\end{minipage}
}
\caption{Two sequential gates operating on the first three bits and one bit from the rest.}
\label{fig:gate}
\end{figure}
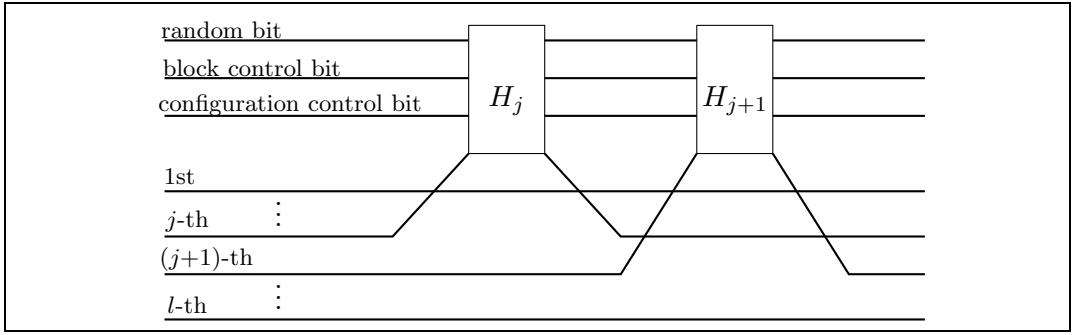

For the above operations, we can use some gates operating on no more than four bits that are the first three bits and one bit from the rest in each time. With $ l $ sequential gates, we can determine whether we are in $ C_j $ or not. Similarly, with $ l $ sequential gates, we can implement the transition from $ C_j $ to $ C_j' $ and, with another $ l $ sequential gates, we can implement the transition from $ C_j $ to $ C_j'' $. Here, using $l$ sequential gates allows us to keep the size of any gate no more than 4 bits as shown in Fig. \ref{fig:gate} (where $H_1,\ldots,H_l$ denote the $l$ sequential gates).

When physically implementing the above circuit $K_{M,x}$, before each block, the circuit will be in a single configuration, and during executing the block, only the part corresponding to this configuration will be active, and thus the circuit will switch to one of the two possible next configurations. After \textit{the decision block}, we will observe the first bit as 1 and 0 with probabilities $ A $ and $ R $, respectively.

Remark that the set of all possible gates which can be used in the above circuit is finite and independent of the input $x$. The only probabilistic gate is a single bit operator implementing a fair coin toss. The rest of gates are deterministic and basically they are controlled operators with maximum dimension of $ 16 $. 

Before continuing with the quantum part, we make further simplifications on $K_{M,x}$. As 2-bit AND and OR gates\footnote{We assume that these gates are represented by $ 4 \times 4 $ matrices.} and 1-bit NOT gate form a universal gate set for classical circuits, each deterministic gate (operating at most 4 bits) can be replaced by some finite numbers of NOT, AND, OR, and some 1-bit resetting gates with help of a few extra auxiliary bits used for intermediate calculations, which are appended to the bottom part of the circuit. Let $ G = \{ G_0,G_1,\ldots,G_t\} $ be the new set of our gates, where $ G_0 $ implements the fair coin by outputting the values 0 and 1 with equal probability, and the values are used by the deterministic gates whenever it is needed.

We denote the simplified circuit as $ K'_{M,x} $ or shortly as $ K' $. Let $ l' $ be the width of $ K' $ (note that $l'= l + O(1) =  O(\log n)$ ). Thus, we have $ K' $ such that the probability of observing 1 (resp.~0) on the first bit is $ A $ (resp.~$ R $). 

\subsection{QTM part}\label{sub:part2}

In this subsection, we give a logarithmic-space postselecting QTM that simulates the computation of $ K' $ in a {\em coherent} manner, as described in Section~\ref{sec:introduction}. 

A logarithmic-space (postselecting) QTM can trace the computation of $ K' $ on its quantum tape by help of its classical part. Since the circuit $ K' $ is deterministic logarithmic-space constructible, the classical part of the QTM helps to create the parts of $ K' $ on the quantum tape whenever it is needed. Moreover, any mixture of the configurations in $ K' $ is kept in a pure state of $ l' $ qubits (described below). 

The QTM uses $ l'+2 $ active qubits on the quantum tape for tracing $ K' $ on the input. The last two qubits are auxiliary, and the first $l'$ qubits are used to keep the probabilistic state of $ K' $. We consider the quantum tape as a logarithmic-width quantum circuit simulating $ K' $.

For each gate of $ K' $, say $ G_j $, we apply a unitary gate (operator) operating on at most 4 qubits, say $ U_j $. Therefore, we use 4 tape heads on the quantum tape. 

During the simulation, the first $ l' $ qubits are always kept in a superposition and after each unitary operator the last qubit or the last two qubits are always measured. If the outcome 
is $ 0 $ or $ 00 $, then the computation continues. Otherwise, the computation is terminated in the non-postselecting state. 

In the probabilistic circuit $ K' $, $ G_0 $ is applied on the first qubit. For each $ G_0 $, we apply
\[
	U_0 =
	\frac{1}{2}
	\mymatrix{rrrr}{ 
	1 & 1 & 1 & 1 \\
	1 & 1 & -1 & -1 \\
	1 & -1 & 1 & -1 \\
	1 & -1 & -1 & 1
	  }
\]  
on the first and the last qubits, measure the last qubit, and continue if $\ket{0}$ is observed:
$$  
    U_0 \ket{00} = \frac{1}{2} \ket{00} + \frac{1}{2} \ket{01} + \frac{1}{2} \ket{10} + \frac{1}{2} \ket{11} \xrightarrow{ \mbox{ post-selection}} \frac{1}{\sqrt{2}} \ket{0} + \frac{1}{\sqrt{2}} \ket{1}.
$$
$$  
    U_0 \ket{10} = \frac{1}{2} \ket{00} - \frac{1}{2} \ket{01} + \frac{1}{2} \ket{10} - \frac{1}{2} \ket{11} \xrightarrow{ \mbox{ post-selection}} \frac{1}{\sqrt{2}} \ket{0} + \frac{1}{\sqrt{2}} \ket{1}.
$$
Thus, coin-flipping operator can be easily implemented.

For the other operators (including the ones given below), we use the techniques given in \cite{SY11A}. For any $ G_j $ ($ 1 \leq j \leq t $), we apply unitary operator $ U_j $ acting on four qubits. Before applying $ U_j $, the quantum part is in 
\[
	\sum_{a,b \in \{0,1\}} \alpha_{a,b} \ket{ab00},
\] 
since the last two qubits are measured before and any outcome other than $ \ket{00} $ is discarded by entering the non-postselecting state. Thus, only $ 4 \times 4 = 16 $ entries of $ U_j $ affects the above quantum state. We construct $ U_j  $ step by step as follows. These 16 entries are set to the corresponding values from $ G_j $. Thus, the probabilistic state, which is kept in the pure state, can be traced exactly up to some normalization factor.

Without loss of generality, we assume that (by reordering the quantum states) these 16 values are placed in the top left corner. Then, $ U_j $ is of the form
\[
	 \frac{1}{e} \mymatrix{c|c|c|c}{ G_j & G'_j & G''_j & 0 \\ \hline * & * & * & * },
\]
where $ e $ is the normalization factor and all $ G_j $, $ G_j' $, and $ G_j'' $ are $ 4 \times 4 $ matrices. 

The entries of $ G_j' $ are set in order to make the first four rows pairwise orthogonal:
\[
	G_j'=
	\mymatrix{cccc}{
		1 & 0 & 0 & 0 \\
		\gamma_{1,2} & 1 & 0 & 0 \\
		\gamma_{1,3} & \gamma_{2,3} & 1 & 0 \\
		\gamma_{1,4} & \gamma_{ 2,4 } & \gamma_{ 3,4 } & 0
	},
\]
where the values are set column by column. The values of $ \gamma_{1,2} $, $ \gamma_{1,3} $, and $ \gamma_{1,4} $ are set to the appropriate values such that the first row becomes orthogonal to the second, the third, and the fourth ones, respectively. Similarly, we set the values of the second and third columns. Since $ G_j $ is composed by integers, $ G_j' $ is also composed by integers. 

The entries of $ G_j'' $ are set in order to make the first four rows with equal length, say $ e $, which is a square of an integer:
\[
	G_j'' = \mymatrix{cccc}{ \gamma_1 & 0 & 0 & 0 \\ 0 & \gamma_2 & 0 & 0 \\ 0 & 0 & \gamma_3 & 0 \\ 0 & 0 & 0 & \gamma_4 },
\]
where diagonal entries are picked as the square roots of some integers.
Remark that the entries of $ G_j'' $ does not change the pair-wise orthogonality of the first four rows. Moreover, at this point, the first four rows become pair-wise orthonormal (due to normalization factor $e$). One can easily fill up the rest of the matrix with some arbitrary algebraic numbers in order to have a complete unitary matrix. 

Since the set of $ G $ depends on the transitions of the PTM $ M $, each $ U_j $ can be kept in the description of the QTM. 

By using the above quantum operators, we can simulate $ K' $ with exponentially small probability. Only note that, due to normalization factors, the computation is terminated in the non-postselecting state with some probabilities after applying each unitary gate. 

At the end of the simulation of $ K' $, we separate the first qubit from the rest of qubits, each of which is set to $ \ket{0} $. Then, we have this unnormalized quantum state in the first qubit:
\[
	(1-A)\ket{0}+A\ket{1}=\myvector{ 1-A \\ A }.
\]
The operator 
$
	\mymatrix{rr}{ 1/2 & 3/2 \\ 1/2 & -1/2 }
$
maps the above quantum state to
\[
	\ket{\tilde{u}} = \myvector{ \frac{1}{2}+A \\ \\ \frac{1}{2} - A }.
\]
Since this operator can be also implemented with post-selection by using an extra qubit, the new unnormalized quantum state is set to $ \ket{\tilde{u}} $. 

If $ A = 0 $, then the quantum state $ \ket{u} $, that is the normalized version of $\ket{\tilde{u}}$, is identical to $ \ket{+} = \frac{\ket{0}+\ket{1}}{\sqrt{2}} $. If $ A < \frac{1}{2} $, then the quantum state $ \ket{u} $ lies between $ \ket{+} $ and $ \ket{0} $, and thus it is closer to $ \ket{+} $ compare to $ \ket{-} = \frac{\ket{0}-\ket{1}}{\sqrt{2}}  $. If $ A > \frac{1}{2} $, then $ \ket{u} $ lies between $ \ket{0} $ and $ \ket{-} $, and thus it is closer to $ \ket{-} $ compare to $ \ket{+} $. 

The measurement in $ \{ \ket{+},\ket{-} \} $ basis means that  we rotate the quantum state with angle $ \frac{\pi}{4} $ in counter-clockwise direction and then make a measurement in $\{\ket{0},\ket{1}\}$ basis. Thus, observing $ \ket{-} $ ($\ket{+}$) in the former case is equivalent to observing $ \ket{0} $ (resp., $ \ket{1} $) in the latter case.

After making a measurement in $ \{ \ket{+},\ket{-} \} $ basis, we can easily distinguish the cases whether $ A $ is close to 0 or $ A $ is close to 1 with bounded error. In the case of when $ A $ is close to $ \frac{1}{2} $, the probability of observing these basis states can be very close to each other. In Section \ref{sub:part3}, we use a modified version of the trick used by Aaronson \cite{Aar05} to increase the success probability. 
Actually, we will need to use the above QTM $O(n^k)$ times sequentially in logarithmic space. 

\subsection{Executing a series of QTMs}
\label{sub:part3}

Let $ p $ be our integer parameter from the set $ \{ 0,1,\ldots,n^k \} $. 
For each $ p $, we consider a QTM $M[p]$ as follows. 
First, we execute the above QTM in Section~\ref{sub:part2}, 
and then transform $ \ket{\tilde{u}} $  to
\[
	\ket{\tilde{u}_p} =  \myvector{ \frac{1}{2}+A \\ \\ 2^{n^k-p} \mypar{ \frac{1}{2} - A } }
\]  
in ($n^k-p$) iterations. In each iteration, we combine the first qubit with another qubit in state $ \ket{0} $, apply the quantum operator 
\[
    \frac{1}{2}\mymatrix{rrrr}{
        1 & \sqrt{3} & 0 & 0 \\ 
        \sqrt{3} & -1 & 0 & 0 \\
        0 & 0 & 2 & 0 \\
        0 & 0 & 0 & 2
    },
\]
and then the second qubit is measured. If the measurement outcome is $ \ket{0} $, then the computation continues. Otherwise, the computation is terminated by entering the  non-postselecting state. (By induction, we can easily see that $ \ket{\tilde{u}} \xrightarrow{~~n^k-p~~\mbox{steps}~~} \ket{\tilde{u}_p} $.) 
Note that 
for each $ p $, the QTM $ M[p] $ can be done in logarithmic space
as the QTM described in Section~\ref{sub:part2} is done in $O(\log n)$ space,
and the counter for the iteration for creating $ \ket{\tilde{u}} $ needs $O(\log n)$ space as well. 

By substituting $ A = \frac{A'}{2^{n^k}} $, the quantum state $ \ket{\tilde{u}_p} $ can be rewritten as
\[
	\ket{\tilde{u}_p} = \myvector{ \frac{1}{2}+ \frac{A'}{2^{n^k}}  \\ \\ 2^{n^k-p} \mypar{ \frac{1}{2} - \frac{A'}{2^{n^k}} } }
	=
	\myvector{ \frac{1}{2}+ \frac{A'}{2^{n^k}}  \\ \\ 2^{n^k-p} \mypar{ \frac{2^{n^k} - 2A'}{2^{n^k+1}} } }
	=
	\myvector{ \frac{1}{2}+ \frac{A'}{2^{n^k}}  \\ \\ \frac{2^{n^k}-2A'}{2^{p+1}} }
	.
\]
It is easy to see that 
\begin{itemize}
	\item when $ A<\frac{1}{2} $ $\left( A' < \frac{2^{n^k}}{2} \Rightarrow 2A' < 2^{n^k} \right)$, the normalized state $ \ket{u_p} $ of $ \ket{\tilde{u}_p} $ lies in the first quadrant, and thus it is closer to $ \ket{+} $, and 
    \item  when $ A>\frac{1}{2} $ $ \left( A' > \frac{2^{n^k}}{2} \Rightarrow 2A' > 2^{n^k} \right) $, $ \ket{u_p} $ lies in the fourth quadrant, and thus it is closer to $ \ket{-} $.
\end{itemize} 

\begin{figure}[h]
\centering
\fbox{
\begin{minipage}{0.85\textwidth}
\centering
\begin{tikzpicture}
\draw (2,2) circle (2cm);

\draw[thick,<->] (-0.5,2) -- (4.5,2);

\draw[thick,<->] (2,-0.5) -- (2,4.5);

\draw[dashed,<->] (0.58,0.58) -- (3.42,3.42);

\draw[dashed,<->] (0.58,3.42) -- (3.42,0.58);

\node at (3.7,3.6) {\footnotesize $ \ket{+} $};

\node at (3.7,0.4) {\footnotesize $ \ket{-} $};

\draw[thick,-] (2,2) -- (2.485,3.94);
\node at (2.5,4.2) {\footnotesize $ \ket{y} $};

\draw[thick,-] (2,2) -- (2.8,3.84);
\node at (3.1,4.05) {\footnotesize $ \ket{u_{p'}} $};

\draw[thick,-] (2,2) -- (2.485,0.06);
\node at (2.6,-0.15) {\footnotesize $ \ket{y'} $};

\draw[thick,-] (2,2) -- (2.8,0.16);
\node at (3.2,0) {\footnotesize $ \ket{u_{p''}} $};

\end{tikzpicture}
\end{minipage}
}
\caption{The visualization of $\ket{y}$ and $ \ket{u_{p'} }$ when $ A < \frac{1}{2} $, and $ \ket{y'} $  and $ \ket{u_{p''}} $ when $ A > \frac{1}{2} $. }
\label{fig:states}
\end{figure}
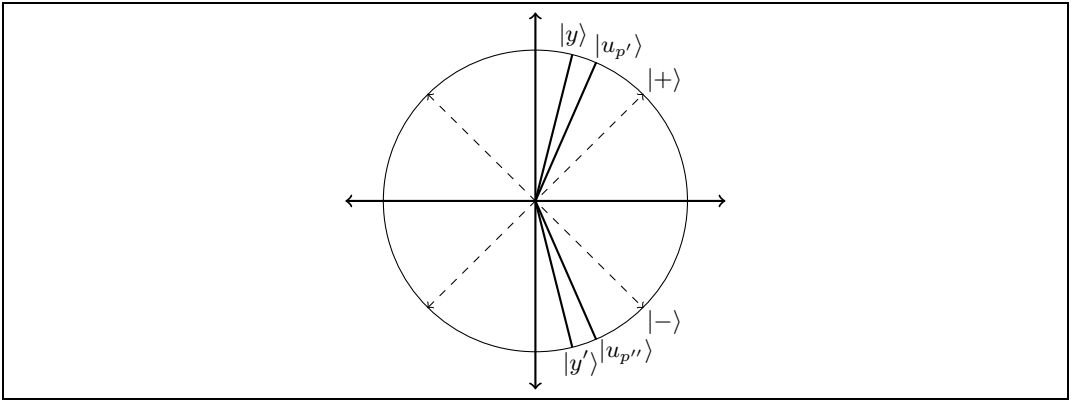

Case $ A < \frac{1}{2} $: As $A\leq \frac{1}{2}-\frac{1}{2^{n^k}}$ (recall that $A$ is the accepting probability of the PTM $M$ on input $x$ that halts in $n^k$ steps), 
\[
\frac{2^{n^k}-2A'}{2^{p+1}}\geq \frac{2}{2^{p+1}}.
\]
Thus, there exists a value of $ p $, say $ p' $, such that 
\[
	\frac{2^{n^k}-2A'}{2^{p'+1}} \in [1,2].
\]
Then, since
$\frac{1}{2}\leq \frac{1}{2}+\frac{A'}{2^{n^k}}$ and $2\geq \frac{2^{n^k}-2A'}{2^{p'+1}}$, 
the quantum state $ \ket{y} = \frac{2}{\sqrt{17}} \myvector{1/2 \\ 2} $ lies between $ \ket{1} $ and  $ \ket{u_{p'}} $, 
and since $\frac{1}{2}\leq \frac{1}{2}+\frac{A'}{2^{n^k}}<1$ and $2\geq \frac{2^{n^k}-2A'}{2^{p'+1}} \geq 1$, 
 $ \ket{u_{p'}} $ lies between $ \ket{y} $ and $ \ket{+} $ (see Fig. \ref{fig:states}). Thus, the probability of observing $ \ket{+} $ after measuring $ \ket{u_{p'}} $ in $ \{\ket{+},\ket{-} \} $ basis is always greater than 
\[
	\frac{25}{34} > \frac{7}{10}
\]
since $ |\braket{y}{+}|^2 = \frac{25}{34} $.

Case $ A>\frac{1}{2} $: The case is similar to the previous case. There exists a value of $p$, say $p''$, such that
\[
	\frac{2^{n^k}-2A'}{2^{p''+1}} \in [-2,-1].
\]
Then, the quantum state $ \ket{y'} = \frac{2}{\sqrt{17}} \myvector{ 1/2 \\ -2 } $ lies between $ -\ket{1} $ and $ \ket{u_{p''}} $ and $\ket{u_{p''}}$ lies between $\ket{y'}$ and $\ket{-}$ (see Fig. \ref{fig:states}). Thus the probability of observing $ \ket{u_{p''}} $ when measuring in $ \{ \ket{+},\ket{-} \} $ basis is always greater than 
\[
	\frac{25}{34} > \frac{7}{10}.
\]

Now the overall quantum algorithm is as follows:
\begin{enumerate}
	\item Prepare counter $C$ to $0$. For each $ p \in \{ 0,1,\ldots,n^k-1 \} $, the following steps are implemented.
	\begin{enumerate}
	    \item We execute the above QTM $M[p]$, and make the measurement at the end in $ \{ \ket{+},\ket{-} \} $ basis. (Note that the execution can be discarded by entering the non-postselecting state in the procedure of Section~\ref{sub:part2}.) 
	    \item If the measurement result corresponds to $\ket{+}$, then we reset the quantum register to all $|0\rangle$ (note that this is possible using the classical control since all the non-$\ket{0}$ qubits are induced only by  post-selection, and thus we know what states they are in), and add $+1$ to $C$.
	    \item If the measurement result corresponds to $\ket{-}$, then we reset the quantum register to all $|0\rangle$, and add $-1$ to $C$.
	\end{enumerate}
	\item If $C=n^k$ (namely, we observe $ \ket{+} $ in all executions), then the input is rejected.
	\item If $C=-n^k$ (namely, we observe $ \ket{-} $ in all executions), then the input is accepted.
	\item Otherwise (namely, if we observe the outcomes $ \ket{+} $ and $ \ket{-} $ at least once in some executions), the computation is terminated in the non-postselecting state. 
\end{enumerate}

Note that the overall quantum algorithm is implemented in logarithmic space 
since the counter is clearly implemented in $O(\log n)$ space, and $M[p]$ is also implemented in $O(\log n)$ space, and each iteration of step 1 is done by the reuse of the classical and quantum registers. 

The analysis of the algorithm is as follows:
\begin{itemize}
	\item When $ A < \frac{1}{2} $, the probability of observing $ \ket{+} $ is always greater than $ \ket{-} $ in each execution and at least once it is $ \frac{7}{3} $ times more. Thus, if $ x \notin L $, the probability of observing all $ \ket{+} $'s is at least  $ \frac{7}{3} $ times more than the probability of observing all $ \ket{-} $'s after all executions. 
    \item When $ A > \frac{1}{2} $, the probability of observing $ \ket{-} $ is always greater than $ \ket{+} $ in each execution and at least once it is $ \frac{7}{3} $ times more. Thus, if $ x \in L $, the probability of observing all  $ \ket{-} $'s is at least $ \frac{7}{3} $ times more than the probability of observing all $ \ket{+} $'s after all executions.
\end{itemize}

Therefore, after normalizing the final accepting and rejecting postselecting probabilities, it follows that $ L $ is recognized by a polynomial-time logarithmic-space postselecting QTM with error bound $ \frac{3}{10} $. This completes the proof of Theorem 2. (The error bound can easily be decreased by using the standard probability amplification techniques.)

\subsection{Additional result}
Additionally, we can show that $\sf PostBQL$ is contained in the class of languages recognized by logarithmic space  bounded-error QTMs that halt in {\em expected exponential time}.

\THM{3}{
$ \sf PostBPL \subseteq BPL(exp) $ and $ \sf PostBQL \subseteq BQL(exp) $, where $\sf BPL(exp)$ ($\sf BQL(exp)$) is the class of languages recognized by logarithmic space  bounded-error PTMs (QTMs) that halt in  {expected exponential time}.
}
\begin{proof}
Let $ M $ be a polynomial-time logarithmic-space PostPTM. By restarting the whole computation from the beginning instead of entering the non-postselecting state, we can obtain a logarithmic-space exponential-time PTM $ M' $ from $ M $, i.e., (i) the restarting mechanism does not require any extra space, and, (ii) since $ M $ produces no less than exponentially small halting probability in polynomial time, $ M' $ halts with probability 1 in exponential expected time. Both machines recognize the same language with the same error bound since the restarting and postselecting mechanism can be used interchangeably \cite{YS10B,YS13A}, i.e., the accepting and rejecting probabilities by $ M $ and $ M' $ are the same on every input. Thus, we can conclude that $ \sf PostBPL \subseteq BPL(exp) $. In the same way, we can obtain that $ \sf PostBQL \subseteq BQL(exp) $.
\end{proof}

As $\sf BQL(exp)\subseteq BQL(\infty)\subseteq PQL$ by definition and Watrous showed $\sf PQL=PL$~\cite{Wat03}, our main result ($\sf PL=PostBQL$) leads to the following equivalence among $\sf BQL(exp)$, $\sf BQL(\infty)$ and $\sf PL$.

\CORR{1}{ $ \sf PL = PQL = PostBQL = BQL(exp) = BQL(\infty) $. }\\

We leave open whether $ \sf BPL(exp)$ is contained in $\sf PostBPL $.

\section{Related Results}
\label{sec:consequences}

In this section, we provide several results on logarithmic-space complexity classes with post-selection. The first result is a characterization of $\sf NL$ by logarithmic-space complexity classes. 

\THM{4}{
	$ \sf NL = PostEPL = PostRL $.
}
\begin{proof}
We start with the first equality $ \sf NL=PostEPL $. Let $ L \in \mathsf{NL} $. Since $ \sf NL = coNL $~\cite{Imm88}, $ \overline{L} $ is also in $\sf NL$. Then, there exist polynomial-time logarithmic-space NTMs $ N_1 $ and $ N_2 $ recognizing $ L $ and $ \overline{L} $. Based on $ N_1 $ and $ N_2 $, we can construct a polynomial-time logarithmic-space PostPTM $ M $ such that $ M $ executes $ N_1 $ and $ N_2 $ with equal probability on the given input. Then, $ M $ accepts the input if $ N_1 $ accepts and rejects the input if $ N_2 $ accepts. Any other outcome is discarded by $ M $. Therefore, (i) any $ x \in L $ is accepted with nonzero probability and rejected with zero probability by $ M $, and, (ii) any $ x \in \overline{L} $ is accepted with zero probability and rejected with nonzero probability by $ M $. Thus, $ L $ is recognized by $ M $ with no error, and thus $ L \in \mathsf{PostEPL} $.
    
Let $ L \in \mathsf{PostEPL} $. Then, there exists a polynomial-time logarithmic-space PostPTM $ M $ recognizing $ L $ with no error. Based on $ M $, we can construct a polynomial-time logarithmic-space NTM $ N $ such that $ N $ executes $ M $ on the given input and switches to the rejecting state if $ M $ ends in the non-postselecting halting state. Thus, $ N $ accepts all and only strings in $ L $. Therefore, $ L \in \mathsf{NL} $. 
    
 Now we are done with equality $ \sf NL=PostEPL $. It is trivial that $ \sf PostEPL \subseteq PostRL$. To complete the proof, it is enough to show that $ \sf PostRL \subseteq NL $. If a language is recognized by a polynomial-time logarithmic-space PostPTM $ M $ with one-sided bounded-error, then it is also recognized by a polynomial-time logarithmic-space NTM $ M' $ where $ M' $ is modified from $ M $ such that if $ M $ enters the non-postselelecting state, then $ M' $ enters the rejecting state.
\end{proof}

By using the same argument, we can also obtain the following result on quantum class $ \sf PostEQL $ (note that the first equality comes from $\sf NQL=coC_=L$ \cite{Wat99A,FR21}).

\THM{5}{$ \sf C_=L \cap coC_=L = NQL \cap coNQL = PostEQL $.}
\\

As will be seen below, the relation between $ \sf PostEQL $ and $ \sf PostRQL $ seems different from the relation between their classical counterparts since $ \sf C_{=}L $ and $ \sf coC_{=}L $ may be different classes. Remark that it is also open whether $ \sf NL $ is a proper subset of $ \sf C_=L \cap coC_=L $ or not.

By using the quantum simulation given in Section~\ref{section:main}, we can obtain the following result.

\THM{6}{$ \sf coC_{=}L = PostRQL $.}
\begin{proof}
It is easy to see that $ \sf PostRQL \subseteq NQL $. Let $ L $ be a language in $ \sf PostRQL $ and $ M $ be a polynomial-time logarithmic-space PostQTM recognizing $ L $ with one-sided bounded-error. By changing the transitions to the non-postselecting state of $ M $ to the rejecting state, we can obtain a polynomial-time logarithmic-space NQTM recognizing $ L $, and thus $ \sf PostRQL \subseteq NQL $. Since $ \sf NQL = coC_{=}L $ \cite{Wat99A,FR21}, we obtain $ \sf PostRQL \subseteq coC_{=}L $.

Now we prove the other direction. Let $ L $ be in $ \sf coC_{=}L $. Then there exists a polynomial-time logarithmic-space PTM $ M' $ that accepts any non-member of $ L $ with probability $ \frac{1}{2} $ and any member with probability different from $ \frac{1}{2} $. Let $ x $ be a given input with length $ n $. 
    
We use the simulation given in Section~\ref{section:main}. We make the same assumptions on the PTM $ M' $ except that $ M' $ accepts some string with probability $ \frac{1}{2} $ and $ M' $ never accepts any string with probability in the following interval
\[
    \left( \frac{1}{2} - \frac{1}{2^{n^k}}, \frac{1}{2} + \frac{1}{2^{n^k}} \right) 
\]
for some fixed integer $ k $. This condition is trivial if the running time never exceeds $ n^{k} $, i.e., the total number of probabilistic branches never exceeds $ 2^{n^k} $.  

Then, we construct a polynomial-time logarithmic-space PostQTM as described in Section~\ref{section:main} with the following unnormalized final quantum state:
\[
	\myvector{ 2A - 1 \\ 2^{-n^k} },
\]
where $A$ is the accepting probability of $M'$. 
We measure this qubit and accept (reject) the input, if we observe $ \ket{0} $ ($ \ket{1} $). All the other outcomes are discarded by entering the non-postselecting state.

It is clear that for any non-member of $ L $, $ A $ is always equal to $ \frac{1}{2} $, and thus the QTM accepts the input with zero probability and rejects the input with some non-zero probability. Therefore, any non-member of $ L $ is rejected 
with probability 1.

On the other hand, for any member, the amplitude of $ \ket{0} $ is at least twice of the amplitude of $ \ket{1} $, and thus the accepting probability is at least four times more than the rejecting probability. Thus, any member is accepted 
with probability at least $ \frac{4}{5} $. The success probability can be increased by using the standard probability amplification techniques.  
\end{proof}

\section*{Acknowledgements}
Part of this research was done while Yakary{\i}lmaz was visiting Kyoto University in November 2016 and March 2017. 
Yakary{\i}lmaz was partially supported by ERC Advanced Grant MQC and ERDF project Nr. 1.1.1.5/19/A/005 ``Quantum computers with constant memory''. Le Gall was supported by JSPS KAKENHI grants Nos.~JP19H04066, JP20H05966, JP20H00579, JP20H04139, JP21H04879 and by the MEXT Quantum Leap Flagship Program (MEXT Q-LEAP) grants No.~JPMXS0118067394 and JPMXS0120319794. 
Nishimura was supported by JSPS KAKENHI grants Nos.~JP19H04066, JP20H05966, JP21H04879 and by the MEXT Q-LEAP grants No.~JPMXS0120319794. 

\bibliographystyle{plain}
\bibliography{tcs}

\end{document}